\documentstyle[12pt]{article}

\def\seCtion#1{\section{#1} \setcounter{equation}{0}}
\renewcommand\theequation{\ifnum\value{section}>0{\thesection.
\arabic{equation}}\fi}
\newcommand{\be}{\begin{equation}}
\newcommand{\ee}{\end{equation}}
\newcommand{\bea}{\begin{eqnarray}}
\newcommand{\eea}{\end{eqnarray}}

\begin{document}

\pagestyle{empty}
\begin{flushright}
\mbox{FIUN-CP-99/2}
\end{flushright}
\quad\\

\begin{center}
\Large {\bf Dispersion relations at finite temperature\\
and density for nucleons and pions}
\end{center}

\begin{center}
{\bf R. Hurtado $^1$}\\
{\it {Department of Physics}, University of Wales\\
Singleton Park, Swansea, SA2 8PP, United Kingdom}\\
\vspace{0.3cm}
{\bf J. Morales $^1$ and C. Quimbay
\footnote{Associate researchers of Centro Internacional de F\'{\i}sica,
Santaf\'e de Bogot\'a, D.C., Colombia. E-mails:
johnmo@ciencias.ciencias.unal.edu.co,
carloqui@ciencias.ciencias.unal.edu.co}}\\
{\it {Departamento de F\'{\i}sica}, Universidad Nacional de Colombia\\
Ciudad Universitaria, Santaf\'e de Bogot\'a, D.C., Colombia}\\
\vspace{0.8cm}
November 21, 1999\\
\vspace{0.3cm}
To be published in Heavy Ion Physics
\end{center}

\vspace{0.5cm}

\begin{abstract}
We calculate the nucleonic and pionic dispersion relations
at finite temperature $(T)$ and non-vanishing chemical potentials
$(\mu_f)$ in the  context of an effective chiral theory that describes
the strong and electromagnetic interactions for nucleons and pions.
The dispersion relations are calculated in the broken chiral symmetry
phase, where the nucleons are massive and pions are taken as massless.
The calculation is performed at lowest order in the energy expansion,
working in the framework of the real time formalism of thermal field
theory in the Feynman gauge. These one-loop dispersion relations are
obtained at leading order with respect to $T$ and $\mu_f$. We also
evaluate the effective masses of the quasi-nucleon and quasi-pion
excitations in thermal and chemical conditions as the ones of a neutron
star.

\vspace{0.8cm}


\end{abstract}

\newpage

\pagestyle{plain}


\seCtion{Introduction}

\hspace{3.0mm}
Effective chiral theories have become a major conceptual and analytical
tool in particle physics driven by the need of a theory to describe the
low--energy phenomenology of QCD. The foundations were formulated
originally by Weinberg \cite{wein} to characterise the most general
S-matrix elements for soft pion interactions and later it was further
developed by Gasser and Leutwyler \cite{Gass}.  Effective chiral
theories have shown to be an adequate framework to treat low--energy
phenomenology \cite{Dono}-\cite{ecker}, as they reproduce, at lowest
order in the chiral expansion, the most important results from current
algebras including the low--energy theorems, and at next-to-leading
order, they give precise corrections to these results \cite{Dono}. They
have been widely applied to different problems as meson--meson,
meson--baryon, photon--photon, photon--meson and photon--baryon
scattering, photoproduction processes and rare kaon decays
\cite{wise,mosel}.

The propagation properties of relativistic particles in plasmas at
finite temperature is also a subject of increasing interest. It is well
known that the interaction of a particle with a plasma in thermal
equilibrium at temperature $T$ modifies the Dispersion Relations (DR)
with respect to the zero temperature situation. This phenomenon has been
extensively investigated for the non-dense plasma case
\cite{kalas}-\cite{rio}, i.e. when the chemical potential $(\mu_f)$
associated to the fermions of the thermal plasma is equal to zero:
$\mu_f = 0$ and $T \not = 0$. In this case the Fermionic Dispersion
Relations (FDR) have been studied for massless fermions in
\cite{kalas}-\cite{gat} and massive fermions in \cite{pis}-\cite{rio}.
The FDR describe the propagation of the fermionic excitations of the
plasma (quasi-fermions and quasi-holes) through the thermal background.
These excitations are originated in the collective behaviour of the
plasma system at low momentum.

On the other hand, DR describing the propagation of the fermionic
excitations of a dense plasma at finite temperature can be found in
literature \cite{lev}-\cite{kal}. For the dense plasma case at finite
temperature, i.e. $\mu_f \not = 0$ and $T \not = 0$, the FDR have been
calculated both for massless fermions in \cite{lev}-\cite{moral} and for
massive fermions in \cite{kal}. These FDR have been calculated in the
context of realistic physical models, as for instance, the Minimal
Standard Model \cite{qui,moral}.

In the present work we calculate the DR for quasi--nucleons and
quasi--pions propagating in a plasma at finite temperature and
non--vanishing chemical potentials. The calculation is performed for a
$SU(2)_L \times SU(2)_R$ effective chiral Lagrangian with the chiral
symmetry broken into $SU(2)_{L+R}$. This Lagrangian, which we introduce
in section 2, describe the strong and electromagnetic interactions of
massive nucleons and massless pions. The calculation is performed using
the real time formalism of the thermal field theory
\cite{dol}-\cite{land} in the Feynman gauge. The one--loop DR are
calculated at lowest order in the energy expansion and obtained taking
the $T^2$ and $\mu_f^2$ terms from the self--energy, as shown in
section 3. As an application of the DR obtained, we evaluate the
effective masses of the quasi--nucleon and quasi--pion excitations
taking the following values: $T=$150 MeV, $\mu_{p}$=100 MeV and
$\mu_{n}=2\mu_{p}$, being $\mu_{p} (\mu_{n})$ the chemical potential
for protons (neutrons) \cite{culo}. This evaluation is shown in section
4, as well as the discussion of the main results and conclusions.


\seCtion{Effective chiral Lagrangian at leading order in the energy
expansion}

\hspace{3.0mm}
Effective chiral theories are founded in the existence of an energy scale
$\Lambda _\chi$ at which chiral symmetry $SU(N_f)_L \times SU(N_f)_R$,
with $N_f$ the number of flavours, breaks into $SU(N_f)_{L+R}$ leading
to $N_f^2 -1$ Goldstone bosons associated to the $N_f$ broken generators.
These Goldstone bosons are identified with the meson ground state octet
for $N_f=3$, and with the triplet of pions \cite{Gass,ecker} in the case
of $N_f=2$. The chiral symmetry of the Lagrangian is broken through the
introduction of an explicit mass term for the nucleons.

A general form for a Lagrangian with $SU(2)_{L+R}$ symmetry describing
the strong and electromagnetic interactions for massive nucleons and
massless pions is \cite{vol,chang}:
\begin{equation}
{\cal L}=\frac{F_{\pi}^2}{4} Tr \left[ D_\mu \Sigma D^\mu \Sigma^{\dag}
\right] + {\cal L}_{\pi N}-\frac{1}{4} F_{\mu \nu} F^{\mu \nu},
\label{Lag}
\end{equation}
where 
\begin{eqnarray}
{\cal L}_{\pi N}&=&\bar N i \gamma^\mu \partial_\mu N - ie \bar N
\gamma^{\mu} A_\mu \left( \frac{1+ \tau_3}{2} \right)N + M \bar N N
\nonumber \\
&+&iM g_A \bar N \gamma_5 \frac{\tau \cdot \pi}{2F_\pi}N + M g^2_A
\bar N \left( \frac{\tau \cdot \pi }{2F_\pi}\right)^2N + ...,
\end{eqnarray}
with 
\begin{eqnarray}
F_{\mu \nu} &=& \partial_\mu A_\nu- \partial_\nu A_\mu, \\
\Sigma &=& e^{i \frac{\tau \cdot \pi}{F_\pi}},
\end{eqnarray}
where the covariant derivative and electromagnetic charge are defined as
\begin{eqnarray}
D_\mu \Sigma &=& \partial_\mu \Sigma + ie A_\mu [Q,\Sigma], \\
Q &=& \left( 
\begin{array}{cc}
\frac{2}{3} & 0 \\ 
0 & \frac{-1}{3}
\end{array}
\right).
\end{eqnarray}
Here $\pi$, $N$ and $A_\mu$ represent the pion, nucleon and
electromagnetic fields, $F_\pi$=93 MeV is the pion decay constant,
$e$ is the electromagnetic coupling constant, $g_A=1.26$ is the axial
coupling constant, and $M$ is the average nucleon mass.


\seCtion{Dispersion relations for nucleons and pions}

\hspace{3.0mm}
In this section we calculate the DR for nucleons and pions in the
framework of the Lagrangian given by (\ref{Lag}). We consider the
propagation of the nucleonic and pionic excitations in a dense thermal
plasma constituted by protons, neutrons, charged pions, neutral pions
and photons, being this plasma characterised by $\mu_{f_i} \not = 0$,
where $f_i$ represents the different fermion species. The calculation
is performed in the real time formalism of the thermal field theory in
the Feynman gauge. The real part of the nucleonic and pionic
self-energies are evaluated at lowest order in the energy expansion and
at one-loop order $(g_A/F_\pi)^2$, considering only the leading
contributions in $T$ and $\mu_{f}$.

The Feynman rules for the vertices at finite temperature and density
(Fig. 1) are the same as those at $T=0$ and $\mu_f=0$, while the
propagators in the Feynman gauge for photons $D_{\mu \nu}(p)$, pions
$D(p)$ and massive nucleons $S(p)$ are \cite{kobes}:
\begin{eqnarray}
D_{\mu \nu}(p) &=& -g_{\mu \nu} \left[ \frac{1}{p^2+i\epsilon} -i
\Gamma_b(p) \right],  \label{bp} \\
D(p) &=& \frac{1}{p^2+i\epsilon}-i{\Gamma}_b(p),  \label{ep} \\
S(p) &=& \frac{p{\hspace{-1.9mm}\slash}}{p^2+m_f^2+i \epsilon}+
i p{\hspace{-1.9mm} \slash}{\Gamma}_f(p),  \label{fp}
\end{eqnarray}
where $p$ is the particle four-momentum and the plasma temperature $T$
is introduced through the functions $\Gamma_b(p)$ and $\Gamma_f(p)$,
which are given by
\begin{eqnarray}
\Gamma_b (p)= 2\pi \delta(p^2)n_b (p),  \label{db} \\
\Gamma_f (p)= 2\pi \delta(p^2)n_f (p),  \label{df}
\end{eqnarray}
with 
\begin{eqnarray}
n_b (p) &=& \frac{1}{e^{(p\cdot u)/T}-1},  \label{nb} \\
n_f(p) &=& \theta(p\cdot u)n_{f}^{-}(p)+\theta(-p\cdot u)n_{f}^{+}(p),
\label{nf}
\end{eqnarray}
being $n_b(p)$ the Bose--Einstein distribution function, and the
Fermi--Dirac distribution functions for fermions ($n_{f}^{-}(p)$) and
anti-fermions ($n_{f}^{+}(p)$) are:
\begin{eqnarray}
n_{f}^{\mp}(p)= \frac{1}{e^{(p\cdot u \mp \mu_f)/T}+1}.
\end{eqnarray}
In the distribution functions $(\ref{nb})$ and $(\ref{nf})$,
$u^{\alpha}$ is the four--velocity of the CoM frame of the plasma, with
$u^\alpha u_\alpha =1$.

The broken chiral symmetry, with hadronic non zero densities, has been
predicted to be restored at about the same temperature at which
deconfinement sets in. This temperature is of the order of 200 MeV as it
is indicated by lattice calculations \cite{kogut}.

\subsection{Nucleonic Dispersion Relation}

\hspace{3.0mm}
Using the Feynman diagrams given in Fig. (2), we calculate the FDR for
quasi--protons and quasi--neutrons.

In order to apply a similar procedure to that followed in
\cite{wel,qui,moral}, we first consider the hypothetical case of massless
nucleons. In this case, we obtain two solutions: one describing
the propagation of quasi-fermions
\begin{equation}
w(k) = M_{p,n}+\frac{k}{3}+\frac{k^2}{3M_{p,n}}+{\cal O}(k^3),
\label{dr1}
\end{equation}
and another one describing the propagation of quasi-holes
\begin{equation}
w(k) = M_{p,n}-\frac{k}{3}+\frac{k^2}{3M_{p,n}}+{\cal O}(k^3).
\label{dr2}
\end{equation}
We observe that if $k=0$, $w(k)= M_{p,n}$. Then $M_p (M_n)$ can be
interpreted as the effective mass of the quasi-protons (quasi-neutrons),
and their expressions are:
\begin{eqnarray}
M_p^2=\left(\frac{3g_A^2 M^2}{64F_{\pi}^2}+\frac{e^2}{8}\right)T^2+
\frac{g_A^2 M^2}{32\pi^2F_{\pi}^2} \left(\mu_n^2+
\frac{\mu_p^2}{2}\right )+\frac{e^2\mu_p^2}{8\pi^2}  \label{Mp}
\end{eqnarray}
and
\begin{eqnarray}
M_n^2=\frac{3g_A^2 M^2}{64F_{\pi}^2}T^2+ \frac{g_A^2 M^2}
{32 \pi^2 F_{\pi}^2}\left(\frac{\mu_n^2}{2}+ \mu_p^2\right). \label{Mn}
\end{eqnarray}
For the limit $k>>M_{p,n}$ the FDR are:
\begin{eqnarray}
w(k) &=& k+\frac{M_{p,n}^2}{k}-\frac{M_{p,n}^4}{2k^3}Log(\frac{2k^2}
{M_{p,n}^2})+....
\label{dr3} \\
w(k) &=& k+2ke^{-2k^2/M_{p,n}^2}+....  \label{dr4}
\end{eqnarray}
For very high momentum the relations $(\ref{dr3})$ and $(\ref{dr4})$
become the ordinary DR for a massless fermion propagating in the vacuum,
i.e. $w(k)=k$.

It is easy to demonstrate that $\mu_f=\int^\infty_0 dp
\left[n_f^-(p)-n_f^+(p)\right]$. This result shows that the chemical
potential is associated with the difference between the number of
nucleons over anti--nucleons. The later means that if a dense thermal
plasma is characterised by $\mu_f > 0$, then the plasma presents an
excess of nucleons over anti-nucleons.

Now we consider the more realistic case where nucleons are massive. In
this case the FDR are valid for $T < T_c$, where $T_c \sim 200$ MeV is
the critical temperature of the chiral phase transition in non--zero
hadronic density \cite{kogut}. We observe that $m_{p,n} > M_{p,n}$,
where $m_p (m_n)$ is the rest mass
of the proton (neutron) and $M_{p,n}$ are given by (\ref{Mp}) and
(\ref{Mn}). In the limit $m_{p,n}^2 >> M_{p,n}^2$ the FDR become
\cite{alt}:
\begin{equation}
w(k)^2= k^2 + m_{p,n}^2 + M_{p,n}^2.\label{wpn}
\end{equation}

Starting from relation (\ref{wpn}) and equations (\ref{Mp}), (\ref{Mn}),
we obtain a general expression for the nucleon effective mass splitting
$\Delta M_N^2$:
\begin{eqnarray}
\Delta M_N^2= m_p^2 - m_n^2 +\frac{e^2 T^2}{8}+ \frac{1}{8\pi^2}
\left(\frac{g_A^2 M^2 \mu_n^2}{8 F_{\pi}^2} - \frac{g_A^2 M^2 \mu_p^2}
{8 F_{\pi}^2}+e^2\mu_p^2 \right). \label{deltam}
\end{eqnarray}

\subsection{Pionic Dispersion Relation}

\hspace{3.0mm}
Using the Feynman rules given in Fig. (1), we obtain the following DR for
quasi-pions:
\begin{equation}
w(k)^2 = k^2+M_{\pi^{\pm},\pi^0}^2,
\label{dr2}
\end{equation}
where $M_{\pi^{\pm}}(M_{\pi^0})$ is the effective mass for charged
(neutral) quasi--pions, and their expressions are:
\begin{eqnarray}
M_{\pi^{\pm}}^2=\frac{T^2}{12} \left( \frac{g_A^2 M^2}{F_{\pi}^2}+e^2
\right)+ \frac{g_A^2 M^2}{8 \pi^2 F_{\pi}^2}
\left(\mu_n^2+\mu_p^2\right) \label{Mpicar}
\end{eqnarray}
and
\begin{eqnarray}
M_{\pi^0}^2=\frac{g_A^2 M^2}{8 \pi^2 F_{\pi}^2} \left(
\frac{2 \pi^2 T^2}{3} + \mu_n^2+ \mu_p^2 \right). \label{Mpineu}
\end{eqnarray}

From (\ref{Mpicar}) and (\ref{Mpineu}) we obtain the pion effective
mass splitting $\Delta M_\pi^2$:
\begin{equation}
\Delta M_\pi^2= \frac{e^2 T^2}{12}. \label{msp}
\end{equation}


\section{Results and conclusions}

\hspace{3.0mm}
We now give the results of the calculation for the effective masses of
quasi--nucleons and quasi--pions. We have used the following values
$m_p=$938.271 MeV, $m_n=$939.566 MeV, $M=$938.919 MeV, $T=$150 MeV,
$\mu_p=$100 MeV, $\mu_n=$200 MeV, $e^2=$0.095. The temperature and
chemical potential values are of the order of those in a neutron star
\cite{culo}. The results for the effective masses are:
\begin{eqnarray*}
M_p &=& 1036.5133 \ {\rm MeV} \\
M_n &=& 1033.8394 \ {\rm MeV} \\
M_{\pi^{\pm}} &=& 637.2312 \ {\rm MeV} \\
M_{\pi^0} &=& 637.0914 \ {\rm MeV}
\end{eqnarray*}
where $M_p$, $M_n$, $M_{\pi^{\pm}}$ and $M_{\pi^0}$ are the effective
masses for the proton, neutron, charged pions and the neutral pion,
including the strong and electromagnetic interactions.

The effective mass splitting for nucleons and pions are:
\begin{eqnarray*}
\Delta (M_p-M_n) &=& 2.6740 \ {\rm MeV} \\
\Delta (M_{\pi^{\pm}}-M_{\pi^0}) &=& 0.1398 \ {\rm MeV} \\
\end{eqnarray*}
where $\Delta (M_{\pi^{\pm}}-M_{\pi^0})$ is due exclusively
to the combined electromagnetic interaction and temperature effects,
as shown at (\ref{msp}). For the nucleons, from the total effective mass
splitting $\Delta (M_p-M_n)$, the combined electromagnetic
and temperature contribute is $\Delta_{em} (M_p-M_n)=$ 0.0058 MeV.

In conclusion, temperature effects enter into the effective mass
splitting relations (\ref{deltam}) and (\ref{msp}) exclusively in the
electromagnetic interaction term, which at $T=$0 vanishes. Also, in the
framework of our model we found that, for the chemical potentials and
temperature used, the effective mass on the proton is bigger than the
one of the neutron. Our results should be improved by considering
massive pions and introducing the weak interaction, as well as using a
realistic model for neutron stars, to be presented in short.

\section*{Acknowledgements}
This work was supported by COLCIENCIAS (Colombia), Universidad
Nacional de Colombia and Centro Internacional de F\'{\i}sica. We want
also to thank to Fernando Cristancho by invitation to participate in
the Third Latinamerican Workshop on Nuclear and Heavy Ion Physics,
San Andr\'es, Colombia.


\newpage

\begin{figure}[1]
\let\picnaturalsize=N
\def\picsize{3in}
\def\picfilename{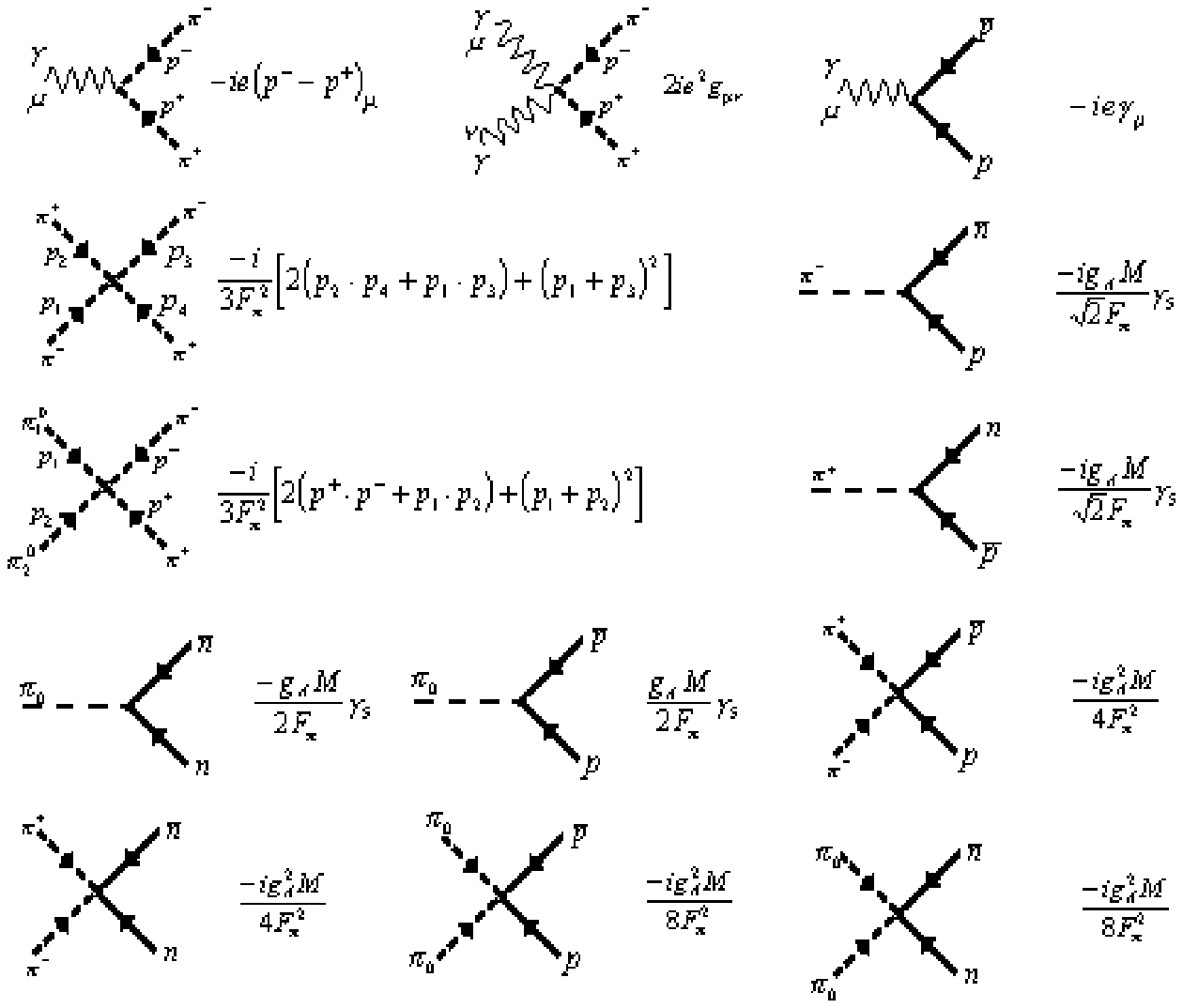}
\caption{Feynman Rules of the ${\cal L}_{\pi N}$.}
\label{Fig.(1)}
\end{figure}

\newpage

\begin{figure}[2]
\let\picnaturalsize=N
\def\picsize{3in}
\def\picfilename{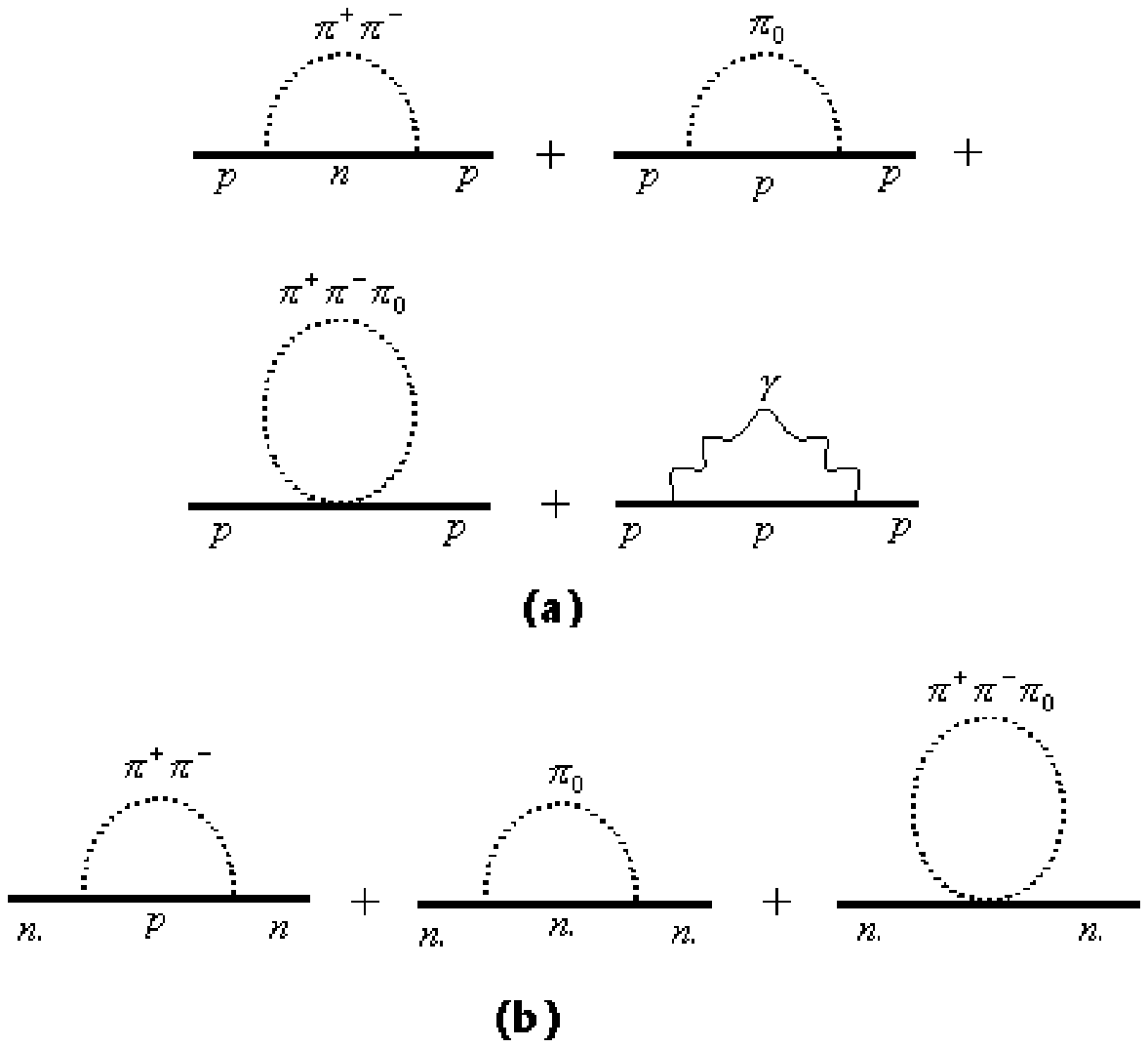}
\caption{Self--energy contributions for the calculation of FDR for:
(a) Protons (b) Neutrons.}
\label{Fig.(2)}
\end{figure}

\end{document}